\RequirePackage{lineno}
\documentclass[twocolumn,aps,prl,showpacs,superscriptaddress]{revtex4}
\usepackage{epsfig,graphics}
\usepackage{graphicx}% Include figure files
\usepackage{dcolumn}% Align table columns on decimal point
\usepackage{bm}% bold math
\usepackage{amsmath}
\usepackage[usenames]{color}
\usepackage{ulem} %% for strike-through

\begin{document}

\title{Dijet asymmetry in Pb+Pb collisions at $\sqrt{s_{_{\rm NN}}}$ = 2.76 TeV within a multiphase transport model}

\author{Guo-Liang Ma}
\affiliation{Shanghai Institute of Applied Physics, Chinese
Academy of Sciences, P.O. Box 800-204, Shanghai 201800, China}

%\date{\today}

\begin{abstract}
Within a multi-phase transport (AMPT) model, dijet asymmetry is studied in Pb+Pb collisions at $\sqrt{s_{_{\rm NN}}}$ = 2.76 TeV. It is found that a large dijet asymmetry ($A_{J}$) is produced by strong interactions between jets and partonic matter. It is demonstrated that hadronization and final-state hadronic rescatterings have little effects on $A_{J}$. The final $A_{J}$ is found to be driven by both initial $A_{J}$ and partonic jet energy loss, which is consistent with an increasing jet energy loss in a hot and strongly interacting partonic medium in more central Pb+Pb collisions.
\end{abstract}


\maketitle

\section{I. Introduction}
\label{sec:intro}
The main goal of high energy heavy-ion collisions is to create a deconfined quark-gluon plasma (QGP) under extreme temperature and density conditions. Jets produced in hard processes is a good probe to search for the formation of partonic matter and investigate its properties, because jets lose energy when they traverse the hot and dense medium~\cite{Wang:1991xy}. The jet quenching phenomenon was first observed through the disappearance of away-side dihadron correlation with high $p_{T}$ leading triggered particle in Au+Au collisions at the Relativistic Heavy Ion Collider (RHIC) ~\cite{Adler:2002tq}. Recently, a large dijet transverse momentum asymmetry, through fully reconstructed jets, has been observed by the ATLAS and CMS experiments at the CERN Large Hadron Collider (LHC), which is consistent with jet partonic energy loss in a QGP medium~\cite{Aad:2010bu,Chatrchyan:2011sx}.  Several theoretical efforts have been made to understand the dijet asymmetry mechanisms so far. Casalderrey-Solana $et~al.$~\cite{CasalderreySolana:2010eh} proposed that frequency collimation can account for the main features seen in the medium-induced dijet asymmetry.  Qin and Muller ~ \cite{Qin:2010mn} reproduced dijet asymmetry with the medium modification of a partonic jet shower traversing in a hot QGP. Young $et~al.$~\cite{Young:2011qx} can fit the dijet asymmetry results with the parton radiative and collisional processes under a MUSIC hydrodynamic background with $\alpha_{s}$= 0.25-0.3. He $et~al.$~\cite{He:2011pd} studied the dijet asymmetry up to $ {\cal O}(\alpha_s^3) $ by including both initial- and final-state nuclear matter effects. Renk  ~\cite{Renk:2012cb} found that the energy dependence of the dijet imbalance is due to the kinematical collimation of jets and the increase in the production probability of quark jets with jet $p_{T}$. In this work, dijet asymmetry is investigated within a multi-phase transport (AMPT) model. A large asymmetry of dijet is found to be produced by strong interactions between jet and partonic matter. In addition, the dijet asymmetry evolution functions are extracted to disclose how the dijet asymmetry evolves from the initial state to the final state through partonic jet energy loss.

\section{II. The AMPT Model}
\label{sec:model}
The AMPT model with string melting scenario~\cite{Lin:2004en},  which has well described many experimental observables~\cite{Lin:2004en,Chen:2006ub, Zhang:2005ni, Ma:2011uma, Ma:2010dv}, is implemented in this work. The AMPT model includes four main stages of high energy heavy-ion collisions: the initial condition, parton cascade, hadronization, and hadronic rescatterings. The initial condition, which includes the spatial and momentum distributions of minijet partons and soft string excitations, is obtained from HIJING model~\cite{Wang:1991hta,Gyulassy:1994ew}. Next it starts the parton evolution with a quark and anti-quark plasma from the melting of strings. The parton cascade process is simulated by Zhang's parton cascade (ZPC) model~\cite{Zhang:1997ej}, where the partonic cross section is controlled by the value of strong coupling constant and the Debye screening mass. Though currently only including elastic parton collisions, the AMPT model can still characterize $\gamma$-jet imbalance at LHC~\cite{Ma:2013bia} with quality similar to that of a linearised boltzmann transport model, which includes both collisional and radiative jet energy loss~\cite{Wang:2013cia}. The AMPT model recombines partons via a simple coalescence model to produce hadrons when the partons freeze-out. The dynamics of the subsequent hadronic rescatterings is then described by a relativistic transport (ART) model~\cite{Li:1995pra}.  In this work, the AMPT model with the newly fitted parameters is used to simulate Pb+Pb collisions at $\sqrt{s_{_{\rm NN}}}$ = 2.76 TeV, which has shown good descriptions for many experimental observables at LHC energy, such as pseudorapidity and $p_{T}$ distributions~\cite{Xu:2011fi} and harmonic flows~\cite{Xu:2011fe, Xu:2011jm}. Two sets of Pb+Pb simulations are performed by setting the partonic interaction cross sections as 1.5 and 0 mb, which correspond to two different physical scenarios for partonic + hadronic interactions and hadronic interactions only, respectively.

\section{III. Jet Reconstruction}
\label{sec:jetrec}

Because the dijet production cross section with large transverse momentum is very small, the dijet production with $p_{T}\sim$ 120 GeV/$c$ is triggered in order to increase the simulation efficiency. Several hard QCD processes are taken into account for the initial dijet production with the jet triggering technique in the HIJING model~\cite{Wang:1991hta,Gyulassy:1994ew}, which includes $q_{1}+q_{2}\rightarrow q_{1}+q_{2}$, $q_{1}+\bar{q_{1}}\rightarrow q_{2}+\bar{q_{2}}$, $q+\bar{q}\rightarrow g+g$, $q+g\rightarrow q+g$, $g+g\rightarrow q+\bar{q}$, and $g+g\rightarrow g+g$, with consideration of initial- and final-state radiation corrections. An anti-$k_{t}$ algorithm from the standard Fastjet package is used to reconstruct full jets~\cite{Cacciari:2011ma}. A pseudorapidity strip of width $\Delta\eta$=1.0 centered on the jet position, with two highest-energy jets excluded, is used to estimate the background (``average energy per jet area"), which is subtracted from the reconstructed jet energy in Pb+Pb collisions. The kinetic cuts are chosen as in the CMS experiment. The jet cone size is set to be 0.5 (R=0.5). The transverse momentum of leading jet is required to be larger than 120 GeV/$c$ ($p_{T,1} >$ 120 GeV/$c$), while that of subleading jet is required to be larger than 50 GeV/$c$ ($p_{T,2} >$ 50 GeV/$c$). The azimuthal angle between leading and subleading jets is larger than $2\pi/3$ ($\Delta\phi_{1,2} > 2\pi/3$), where the subscripts 1 and 2 refer to the leading jet and subleading jet respectively.  Only jets within a mid-rapidity range of $|\eta_{1,2}|<2$ are considered for this analysis. 

\section{IV. Results and Discussions}
\label{sec:results}

\begin{figure}
\includegraphics[scale=0.5]{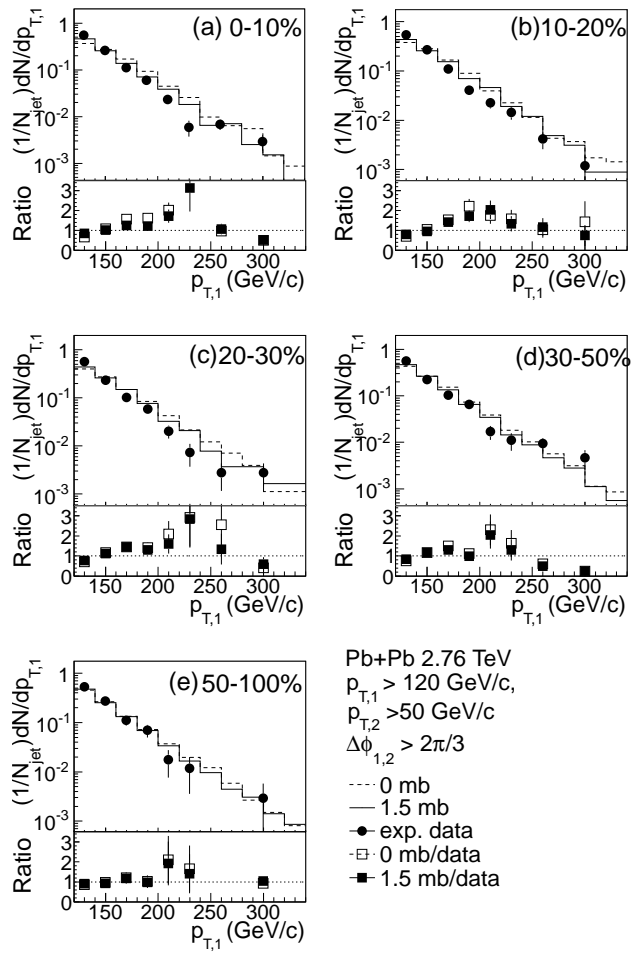}
\caption{Leading jet $p_{T}$ distributions for dijet events with subleading jets of $p_{T,2} >$ 50 GeV/$c$ in Pb+Pb 2.76-TeV collisions for different centrality bins: (a) 0-10\%, (b) 10-20\%, (c ) 20-30\%, (d) 30-50\% and (e) 50-100\%, where the solid (1.5 mb) and dash (0 mb) histograms represent  the AMPT results with partonic+hadronic interactions and hadronic interactions only respectively, while the solid circles represent the data from the CMS experiment~\cite{Chatrchyan:2011sx}. The lower part in each panel depicts the ratios of AMPT results to experimental data.}
 \label{fig-ptdis}
\end{figure}

\begin{figure}
\includegraphics[scale=0.5]{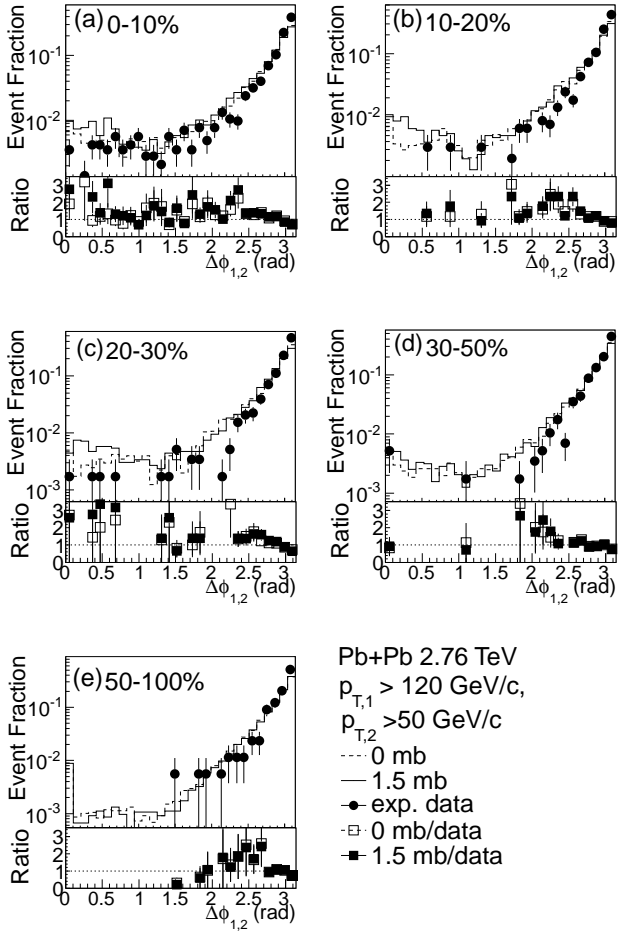}
\caption{$\Delta\phi_{1,2}$ distributions for leading jets of $p_{T,2} >$ 120 GeV/$c$ with subleading jets of $p_{T,2} >$ 50 GeV/$c$ in Pb+Pb 2.76-TeV collisions for different centrality bins: (a) 0-10\%, (b) 10-20\%, (c ) 20-30\%, (d) 30-50\% and (e) 50-100\%, where the solid (1.5 mb) and dash (0 mb) histograms represent  the AMPT results with partonic+hadronic and hadronic interactions only respectively, while the solid circles represent the data from the CMS experiment~\cite{Chatrchyan:2011sx}. The lower part in each panel depicts the ratios of AMPT results to experimental data.
}
 \label{fig-deltaphidis}
\end{figure}

Figure~\ref{fig-ptdis} shows the leading jet $p_{T}$ distributions in Pb+Pb 2.76-TeV collisions for different centrality bins. The AMPT simulations with partonic+hadronic interactions (1.5 mb) give spectra a little softer than those with hadronic interactions only (0 mb). Note that the measured leading jet $p_{T}$ distributions have not been corrected for some detector effects such as detector resolution, fluctuations in and out of the jet cone, or underlying event fluctuations~\cite{Chatrchyan:2011sx}. From a quantitative comparison of the ratios of AMPT results to experimental data (lower part in each panel of Figure~\ref{fig-ptdis}), the AMPT model can reproduce the data to a good degree. 

Figure~\ref{fig-deltaphidis} presents dijet $\Delta\phi_{1,2}$ distributions for different centrality bins in Pb+Pb 2.76-TeV collisions. The dijet $\Delta\phi_{1,2}$ distribution does not seems to be sensitive to whether partonic interactions exist or not, both sets of AMPT results are similar and both can basically describe the experimental data.

\begin{figure}
\includegraphics[scale=0.5]{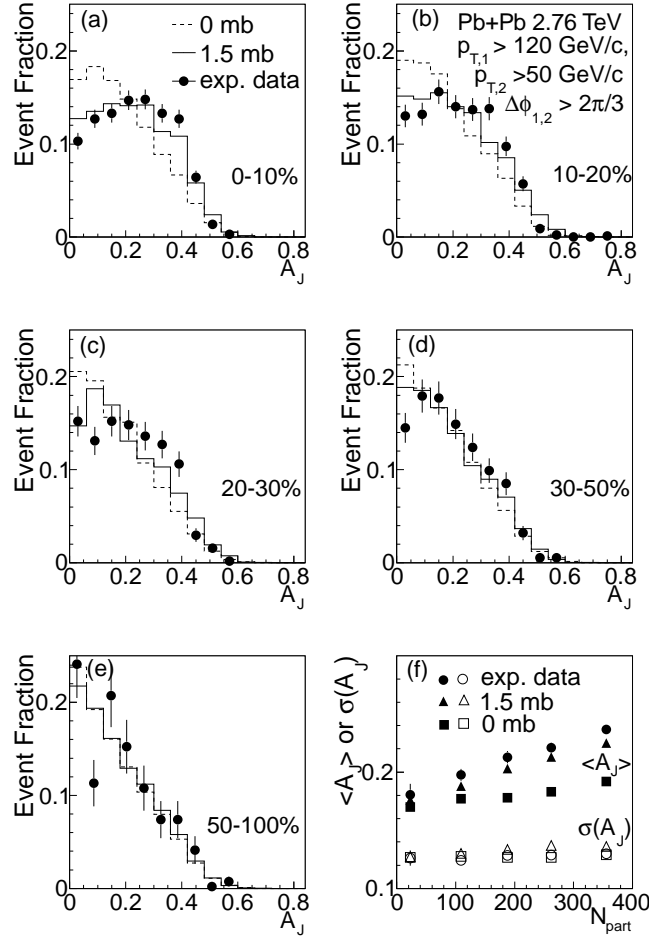}
\caption{Dijet asymmetry ratio $ A_{J}$ distributions for leading jets of $p_{T,2} >$ 120 GeV/$c$ with subleading jets of $p_{T,2} >$ 50 GeV/$c$, and $\Delta\phi_{1,2} > 2\pi/3$ in Pb+Pb 2.76-TeV collisions for different centrality bins: (a) 0-10\%, (b) 10-20\%, (c ) 20-30\%, (d) 30-50\% and (e) 50-100\%, where the solid (1.5 mb) and dash (0 mb) histograms represent  the AMPT results with partonic+hadronic and hadronic interactions only respectively, while the solid circles represent the data from the CMS experiment~\cite{Chatrchyan:2011sx}. (f): the mean values and variances of  $ A_{J}$ distributions as functions of $N_{part}$.
}
\label{fig-AjDis}
\end{figure}

To characterise the transverse momentum balance (or imbalance) of dijet,  an asymmetry ratio is defined as  $A_{J}$=($p_{T,1}-p_{T,2}$)/($p_{T,1}+p_{T,2}$) as LHC experiments did~\cite{Aad:2010bu, Chatrchyan:2011sx}. The dijet asymmetry $A_{J}$ distributions for different centrality bins are shown in Figure~\ref{fig-AjDis} (a)-(e).  For more peripheral collisions, both sets of AMPT results give similar results to describe the data, since the partonic interactions are relatively weak in peripheral collisions. However, it is different for more central collisions where the AMPT results with partonic+hadronic interactions give more asymmetric  $A_{J}$ distribution than those with hadronic interactions only. For instance, for most central centrality bin (0-10\%) in Figure~\ref{fig-AjDis} (a), the AMPT results (1.5 mb) give a much better description than AMPT results (0 mb). Figure~\ref{fig-AjDis} (f) presents the mean values $\left\langle A_{J} \right\rangle$ and variances $\sigma( A_{J})$ of  $ A_{J}$ distributions as functions of number of participant nucleons $N_{part}$. The AMPT results (1.5mb) can well describe the two characteristic quantities for dijet $A_{J}$ distributions simultaneously, however the AMPT results (0 mb) underestimate $\left\langle A_{J} \right\rangle$. This indicates that it is the strong interactions between the jets and the partonic matter that yield the observed large $p_{T}$ imbalance between the two jets.

\begin{figure}
%\hspace{-1.0cm}
\includegraphics[scale=0.45]{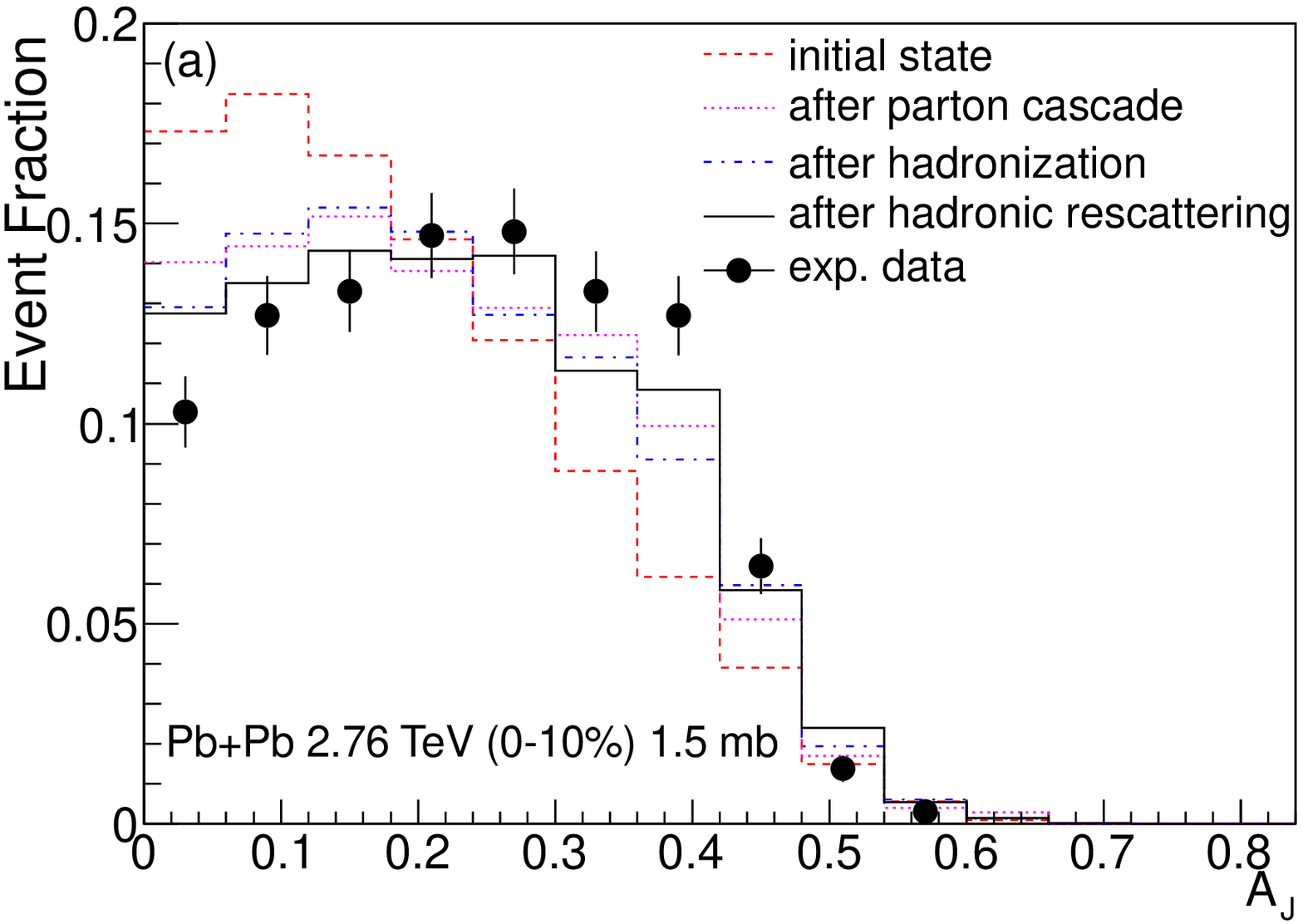}
\includegraphics[scale=0.45]{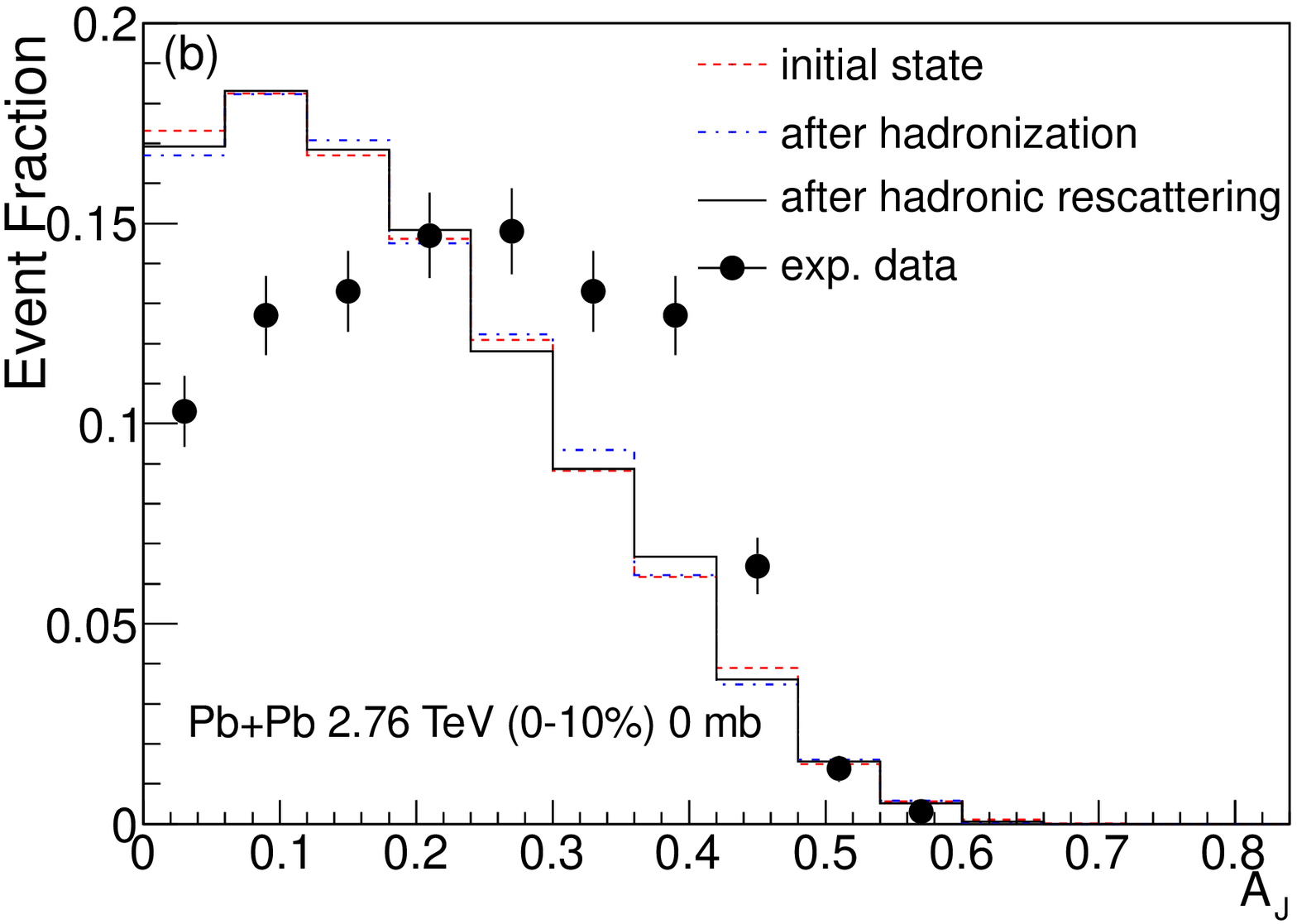}
\caption{(Color online) Dijet asymmetry ratio $ A_{J}$ distributions at different evolution stages for leading jets of $p_{T,2} >$ 120 GeV/$c$ with subleading jets of $p_{T,2} >$ 50 GeV/$c$, and $\Delta\phi_{1,2} > 2\pi/3$ for most central Pb+Pb 2.76-TeV collisions (0-10\%), where panels (a) and (b) represent the AMPT results with partonic+hadronic and hadronic interactions only respectively.}
 \label{fig-stages}
\end{figure}

Heavy-ion collisions are a dynamical evolution including many important stages. It is necessary to compare $A_{J}$ distributions at different stages to learn the effects on dijet asymmetry from different final state processes. Figure~\ref{fig-stages} (a) and (b) give the $A_{J}$ distributions at or after different evolution stages in most central Pb+Pb collisions (0-10\%) for the two sets of AMPT results. In Figure~\ref{fig-stages} (a), the dijet asymmetry ratio $A_{J}$ increases from ``initial state'' to ``after parton cascade'' because jets, especially for subleading jets, lose much energy when they pass through the partonic matter. However, the following hadronization and hadronic rescattering processes do not seem to affect the formed $A_{J}$ distribution any more.  On the other hand, Figure~\ref{fig-stages} (b) shows that it is difficult to produce a visible dijet asymmetry if with hadronic interactions only even for most central collisions.

\begin{figure}
\includegraphics[scale=0.5]{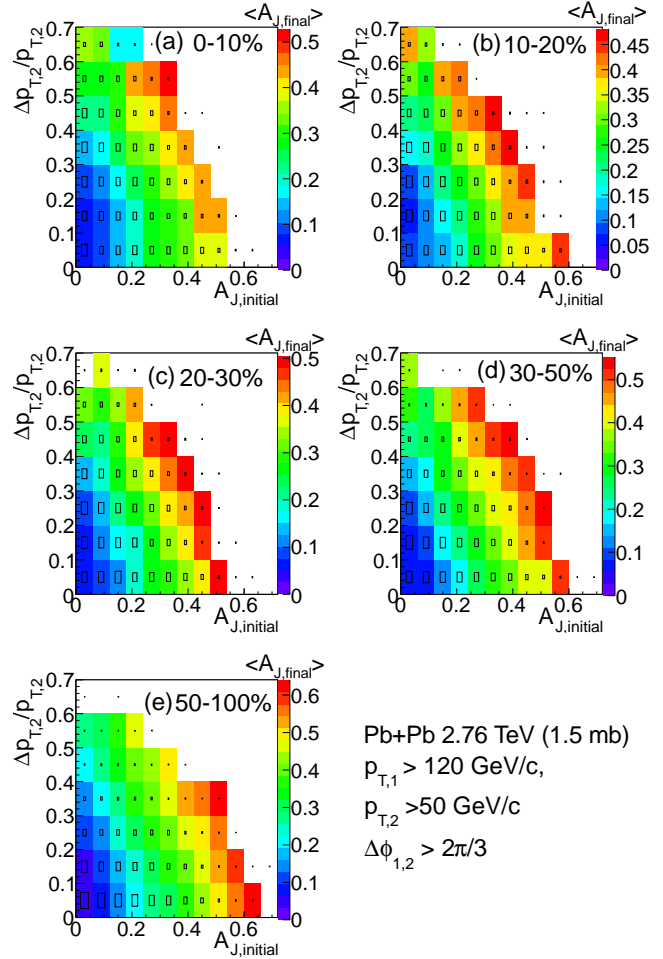}
\caption{(Color online) The AMPT results (1.5 mb) on the reduced dijet $A_{J}$ evolution functions [$ \left\langle A_{J, final} \right\rangle$($ A_{J, initial}$, $\Delta p_{T,2}/p_{T,2}$)] in Pb+Pb 2.76-TeV collisions for different centrality bins: (a) 0-10\%, (b) 10-20\%, (c ) 20-30\%, (d) 30-50\% and (e) 50-100\%, where the size of box in each cell represents the possibilities for dijet events with $ A_{J, initial}$ and $\Delta p_{T,2}/p_{T,2}$.}
 \label{fig-AJevo}
\end{figure}

\begin{figure}
\includegraphics[scale=0.45]{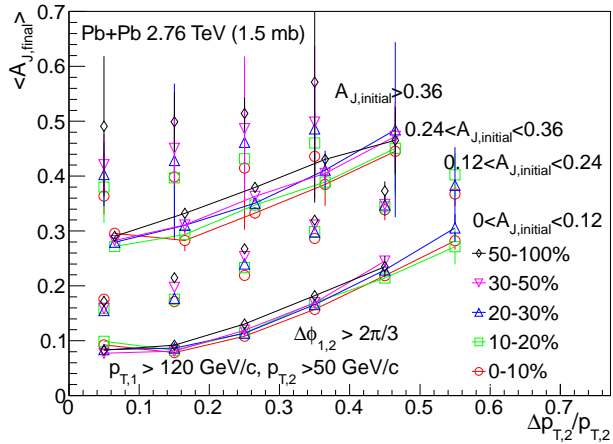}
\caption{(Color online) The AMPT results (1.5 mb) on $ \left\langle A_{J, final} \right\rangle$ as functions of $\Delta p_{T,2}/p_{T,2}$ for given $ A_{J, initial}$ selections in Pb+Pb 2.76-TeV collisions for different centrality bins. Some points are slightly shifted along the $x$ axis for better representation.}
 \label{fig-AJevoprojected}
\end{figure}

\begin{figure}
\includegraphics[scale=0.45]{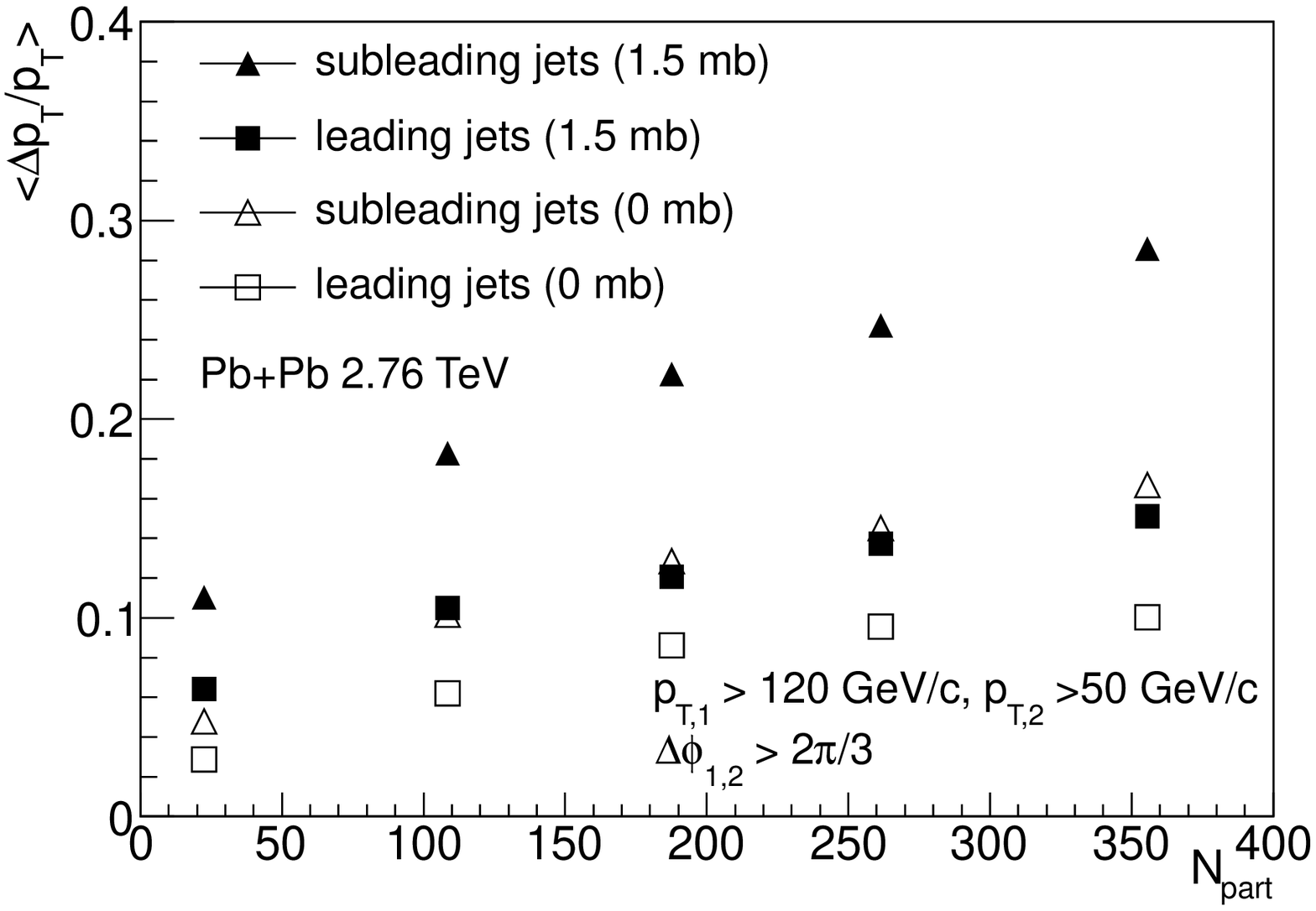}
\caption{Averaged energy loss fractions, $\left\langle\Delta p_{T}/p_{T} \right\rangle$, of leading and subleading jets as functions of $N_{part}$ in Pb+Pb 2.76-TeV collisions, where the solid (1.5 mb) and open (0 mb) symbols represent the AMPT results with partonic+hadronic and hadronic interactions only respectively. }
 \label{fig-eloss}
\end{figure}

To study how the initial $A_{J}$ evolves into the final $A_{J}$, an asymmetry evolution function is defined as the final averaged dijet asymmetry ratio $ \left\langle A_{J, final} \right\rangle$  as a function of the initial dijet asymmetry ratio $A_{J, initial}$ and the energy loss fraction of the leading and subleading jets $\Delta p_{T,1}/p_{T,1}$ and $\Delta p_{T,2}/p_{T,2}$, i.e. $ \left\langle A_{J, final} \right\rangle$ ($ A_{J, initial}$, $\Delta p_{T,1}/p_{T,1}$, $\Delta p_{T,2}/p_{T,2}$), where $\Delta p_{T,1/2}/p_{T,1/2}$=$(p_{T,1/2}^{initial}-p_{T,1/2}^{final})/p_{T,1/2}^{initial}$. Because leading jets lose much less energy than subleading jets due to the dijet surface bias~\cite{Zhang:2007ja}, the dijet asymmetry should be dominantly controlled by the energy loss of subleading jets. The dijet $A_{J}$ evolution function can be reduced to $ \left\langle A_{J, final} \right\rangle$ ($ A_{J, initial}$, $\Delta p_{T,2}/p_{T,2}$), which also is easier for representation. Figure~\ref{fig-AJevo} shows the reduced dijet $A_{J}$ evolution functions for different centrality bins in Pb+Pb 2.76-TeV collisions (1.5 mb), where the color of cell denotes the $ \left\langle A_{J, final} \right\rangle$ and the size of box in each cell represents the possibility of dijet events. It is found that final $A_{J}$ of dijet events is driven by two sources. The first one is the remaining $A_{J}$ part for which dijet keep their initial $A_{J}$ with a small jet energy loss fraction. The second one is the newly formed $A_{J}$ part for which jets lose much more energy to increase their original $A_{J}$, which actually is the dominant source for the large dijet asymmetry in the central collisions. However, the contribution from the second source goes down in more peripheral collisions due to the decrease of partonic interaction strength, which results in less of a change of $A_{J}$ for more peripheral collisions.

Figure~\ref{fig-AJevoprojected} gives the projected dijet $A_{J}$ evolution functions, $ \left\langle A_{J, final} \right\rangle$($\Delta p_{T,2}/p_{T,2}$), for four given $ A_{J, initial}$ selections in Pb+Pb 2.76-TeV collisions for different centrality bins, where four groups of curves correspond to four $ A_{J, initial}$ selections of $0 < A_{J, initial} <0.12$, $0.12 < A_{J, initial} <0.24$, $0.24 < A_{J, initial} <0.36$ and $A_{J, initial} >0.36$ from the bottom up. It shows that the final $A_{J}$ starts to increase from the initial $A_{J}$ and roughly scales with the energy loss fraction of subleading jet for a given initial $A_{J}$ selection. The small violation of scaling is because of the neglect of the energy loss fraction of leading jets in the reduction of the dijet $A_{J}$ evolution function. 

Figure~\ref{fig-eloss} shows that the event averaged energy loss fractions of jets, $\left\langle\Delta p_{T}/p_{T} \right\rangle$, actually increase with $N_{part}$ for both leading and subleading jets for both sets of AMPT simulations. However, the leading jet loses less energy in comparison with subleading jet due to the dijet surface bias. In addition, the interactions between jets and partonic matter indeed can result in jet energy loss fractions larger than those between jets and hadronic matter. The reason the more central collisions show a lower $ \left\langle A_{J, final} \right\rangle$ in Figure~\ref{fig-AJevoprojected} at given $ A_{J, initial}$ and $\Delta p_{T,2}/p_{T,2}$ than peripheral collisions is because the leading jets do lose energy and energy loss increases with $N_{part}$.

\section{V. Summary}
\label{sec:summary}

In summary, dijet transverse momentum asymmetry is investigated within the AMPT model with both partonic and hadronic interactions. It is found that a large dijet asymmetry can be produced by strong interactions between jets and partonic matter. Hadronization and final-state hadronic rescattering have little effects on the dijet asymmetry. It is difficult for hadronic interactions only to reproduce the observed dijet asymmetry, since hadronic interactions make jets lose much less energy than partonic interactions. The $A_{J}$ evolution functions quantitively show that the final dijet asymmetry depends on both the initial dijet asymmetry and the jet energy loss. The large dijet
asymmetry observed in central Pb+Pb collisions indicates a large jet energy loss in the hot and strongly interacting partonic medium created in these collisions.

This work was supported by the NSFC of China under Projects No. 11175232, No. 11035009, No. 11105207, and No. 11220101005; the Knowledge Innovation Program of CAS under Grant No. KJCX2-EW-N01; the Youth Innovation Promotion Association of CAS; and the project sponsored by SRF for ROCS, SEM, and CCNU-QLPL Innovation Fund (Grant No. QLPL2011P01).

\end{document}